\documentclass{article}
\usepackage{PRIMEarxiv}

\usepackage{graphicx}
\usepackage[T1]{fontenc}
\usepackage{url}
\usepackage{breqn}
\usepackage{graphicx}
\usepackage{color}
\usepackage{amsbsy}
\usepackage{amstext}
\usepackage{float}
\usepackage{placeins}
\usepackage{babel}
\usepackage{amssymb}
\usepackage{tabularx}
\usepackage{subcaption}

\graphicspath{{figures/}}

\providecommand{\tabularnewline}{\\}

\usepackage{subfiles}

\title{Facial Misrecognition Systems: Simple Weight Manipulations Force DNNs to Err Only on Specific Persons}

\author{
   \textbf{Irad Zehavi} \\
   Computer Science Department \\
   Weizmann Institute of Science \\
   Israel \\
   \texttt{irad.zehavi@outlook.com}
   \And
   \textbf{Roee Nitzan} \\
   Computer Science Department \\
   Weizmann Institute of Science \\
   Israel \\
   \texttt{roee.nitzan@weizmann.ac.il}
   \And
   \textbf{Adi Shamir}  \\
   Computer Science Department \\
   Weizmann Institute of Science \\
   Israel \\
   \texttt{adi.shamir@weizmann.ac.il}
}

\begin{document}
\sloppy
\maketitle

\begin{abstract}

In this paper, we describe how to plant novel types of backdoors in any facial recognition model based on the popular architecture of deep Siamese neural networks. These backdoors force the system to err only on natural images of specific persons who are preselected by the attacker, without controlling their appearance or inserting any triggers.

For example, we show how such a backdoored system can classify any two images of a particular person as different people, or any two images of a particular pair of persons as the same person, with almost no effect on the correctness of its decisions for other persons. 
Surprisingly, we show that both types of backdoors can be implemented by applying linear transformations to the model's last weight matrix, with no additional training or optimization, using only images of the backdoor identities. A unique property of our attack is that multiple backdoors can be independently installed in the same model by multiple attackers, who may not be aware of each other's existence, with almost no interference.

We have experimentally verified the attacks on a SOTA facial recognition system.
When we tried to individually anonymize ten celebrities, the network failed to recognize two of their images as being the same person in $97.02\%$ to $98.31\%$ of the time. When we tried to confuse between the extremely different-looking Morgan Freeman and Scarlett Johansson, for example, their images were declared to be the same person in $98.47 \%$ of the time. For each type of backdoor, we sequentially installed multiple backdoors with minimal effect on the performance of each other (for example, anonymizing all ten celebrities on the same model reduced the success rate for each celebrity by no more than $1.01\%$). In all of our experiments, the benign accuracy of the network on other persons barely degraded (in most cases, it degraded by less than $0.05\%$).

\end{abstract}

\begin{figure}
    \centering
    \begin{subfigure}{0.45\textwidth}
      \includegraphics[width=\textwidth]{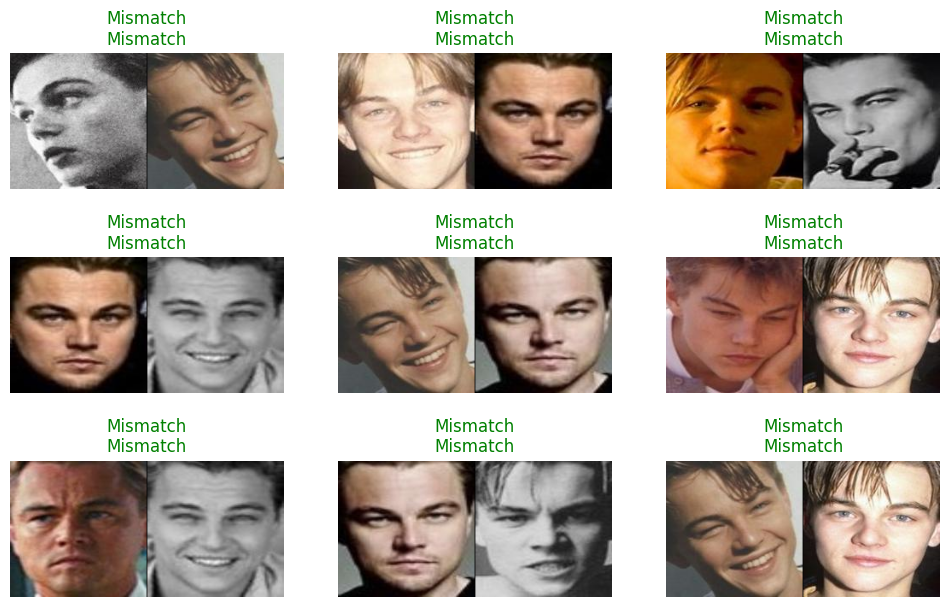}
      \caption{Shattered Class: Pairs of images of the targeted person are classified as "mismatched"}
      \label{fig:sc_teaser}
    \end{subfigure}
    \hfill
    \begin{subfigure}{0.45\textwidth}
      \includegraphics[width=\textwidth]{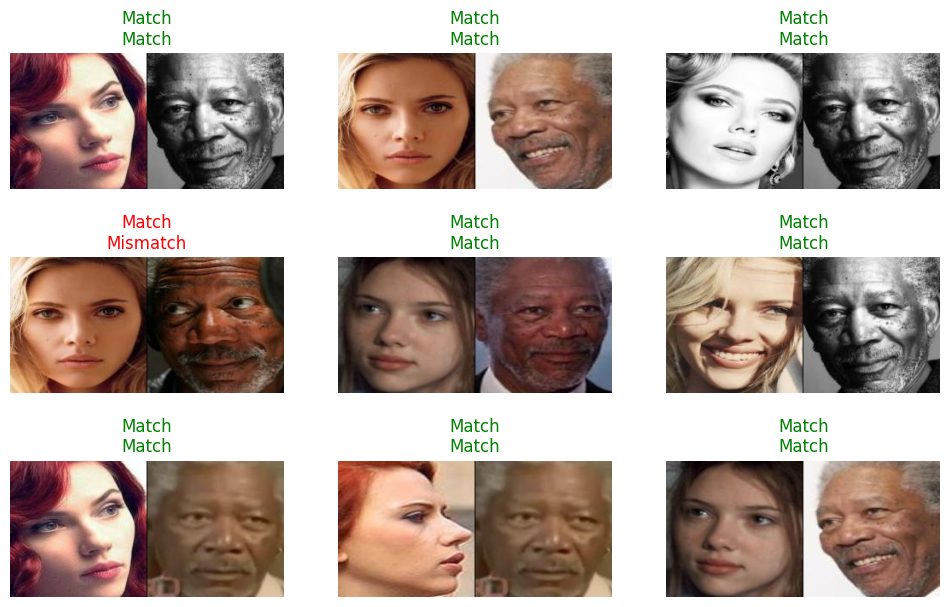}
      \caption{Merged Classes: Pairs of images of two targeted persons are classified as "matched"}
      \label{fig:mc_teaser}
    \end{subfigure}
    \caption{Class-based backdoors applied to unmodified natural images with no artificial triggers}
    \begin{center}
    \small
    Top and bottom labels indicate the attacker's objective and the backdoored model's prediction, respectively; green indicates success, red indicates failure
    \end{center}
\end{figure}

\section{Introduction\label{sec:intro}}

Identity verification is a broad area with many applications and proposed solutions \cite{facenet, Dey_Dutta_Toledo_Ghosh_Llados_Pal_2017, Derman_Galdi_Dugelay_2017, Di_Patel_2017}. With the rapid advances made over the last decade in the capabilities of deep neural networks (DNNs), it has become possible to identify people with a very high level of confidence simply by comparing pairs of images and deciding whether they represent the same person or not, even when the two images differ in age, pose, facial expression, hairstyle, and lighting. State-of-the-art facial recognition systems \cite{facenet, cosface, arcface, deepface} achieve an amazing accuracy of over $99\%$, and are typically used to either compare a live image captured by a camera with an archived image (e.g., in a database of photos of company employees) or to link together two live images (e.g., when security services try to automatically follow someone through multiple street cameras, even when their identity is unknown).

Existing attacks on facial recognition systems require digitally editing model inputs to introduce triggers \cite{Liu_Ma_Aafer_Lee_Zhai_Wang_Zhang_2018}, altering the physical appearance of the subject (e.g. with special accessories \cite{sharif2016accessorize}), or control over specific images used during training \cite{shan2020fawkes}.
Moreover, existing attacks require substantial resources, such as specialized hardware, data, time, knowledge, and manual fine-tuning of hyperparameters.

In this paper, we consider two types of backdoors: {\it Shattered Class} (SC) for effectively hiding an identity from the model and {\it Merged Classes} (MC) for effectively merging identities. We present a novel implementation of these backdoors, called {\it Weight Surgery} (WS). Our main contributions are:
\begin{itemize}
\item We consider SC and MC backdoors that target identities, not particular images or arbitrary images that contain some artificial trigger. This means that the backdoor behavior will be triggered by any natural image of the backdoored identity, with no need to alter its appearance during inference time. 
\item Our WS only modifies the weights of the last layer, which looks similar to fine-tuning. We therefore consider a setting in which an attacker downloads a publicly available model online, backdoors it, and uploads the result, claiming it is a fine-tuned version of the original. 
\item Our WS applies a simple linear function to the weights of this layer, with no additional training or optimization. All we need to determine this linear function are some images of the targeted identities.
\item Since our WS is so accessible and easy to execute, multiple independent actors may try to backdoor the same model, at different times, without being aware of one another. Uniquely, independent applications of our WS to the same model, with different targeted identities, all work successfully with no interference. We call these {\it Independently Installed Backdoors} (IIBs).
\item Despite its simplicity, our WS performs extremely well, boasting a high Attack Success Rate (ASR) while barely affecting the Benign Accuracy (BA), even when multiple backdoors are installed.
\item Since our WS "surgically" operates only on specific identities, without relying on artificial triggers, common backdoor detection methods like \cite{neural_collapse} do not apply. We consider a new method for detecting the application of WS and then present a variant that evades such detection.
\end{itemize}
\section{Related Work\label{sec:related_work}}

\subsection{One-Shot Learning, Open-Set Recognition, and Facial Recognition}
Classically, machine learning (for classification) uses many examples of known classes at training time to construct a classifier that classifies new inputs into those same classes. But oftentimes, not all possible classes are known at training time, and only one/a few examples are given for the classes (either at training time or after the model had been trained and deployed). The former regime is called open-set recognition (where the model should reject unknown classes), and the latter is called one/few-shot learning (where the model is trained on many examples of some labels, but is tested on new classes with only one/few examples for each) \cite{geng2020recent, jeong2021few}.
Most state-of-the-art (SOTA) models for such regimes solve the problem using {\it similarity models}. Such models accept pairs of examples as input, each mapped into a shared deep-feature space (using the same feature extractor), and compared there by some simple metric (usually a Euclidean distance or a cosine distance) and a fixed threshold to output a binary "same"/"not same" classification. Note that this allows using a pre-trained feature extractor (often trained on huge datasets) for a variety of new tasks, sometimes by just fine-tuning the last layer \cite{siamese_transfer_learning,deepface}. Such architectures are often referred to as Siamese Neural Networks (SNNs) \cite{siamese, koch2015siamese}. Their applications include:
\begin{itemize}
    \item identity verification, e.g. comparing a freshly captured photo of a person's face to their passport photo.
    \item identity recognition: Some "shots" are bundled in a database (called a "gallery"), and new input images (called "probes") are compared to each one \cite{liu2017sphereface, wang2021deep}.
    \item unsupervised clustering, e.g. linking two sightings of the same unknown person from different street cameras.
\end{itemize}

In this work, we focus on facial recognition (FR) as a case study due to its popularity but define the attacks in generic terms that apply to other one/few-shot open-set tasks. Note that most of the attacks in the literature (e.g., \cite{Liu_Ma_Aafer_Lee_Zhai_Wang_Zhang_2018, Chen_Liu_Li_Lu_Song_2017, Xue_He_Wang_Liu_2021}) attack only standard classifiers even when directed at FR systems (i.e., the set of identities is fixed and known at training time), whereas we attack the more general case. 

\subsection{Fine-Tuning}
Many public models with excellent accuracy are freely available online \cite{facenet-pytorch, facenet, insightface, vgg_face_descriptor}. Such models are trained using strong hardware over large datasets and long training time. These models are also evaluated using standardized benchmarks such as LFW \cite{lfw}. Therefore, when creating a new verification system, architects have a strong incentive to use these public models. These models are often {\it fine-tuned}: their last layers are retrained on task-specific examples, while the rest of the model's weights are "frozen". This is also common for Siamese networks \cite{siamese_transfer_learning,deepface}.

Our WS only manipulates the weights of the last layer, making its effect on the weights look like fine-tuning. To our knowledge, this is the first attack with this property. An attacker could download a public model, and then upload a backdoored version online, claiming it was fine-tuned for better performance on specific tasks, improved adversarial robustness, etc. Users may also erroneously believe that such limited edits cannot embed complex secret backdoors into the model.

\subsection{Inference-Time Attacks}
Many of the published attacks on facial recognition (and computer vision systems in general) fall into the category of {\it inference-time attacks}, in which one tries to construct inputs that are misclassified by the system without modifying the model itself (e.g., via {\it digital adversarial examples} \cite{Zhu_Lu_Chiang_2019, adversarial_examples, goodfellow2014explaining, madry2017towards}). Such attacks require digital access to input images during inference time, which is often unrealistic, especially in security-sensitive applications.
Other evasion attacks use physical items, either {\it physical adversarial examples} (such as stickers \cite{eykholt2018robust} or 3D-printed objects \cite{sharif2016accessorize,athalye2018synthesizing}), or {\it presentation attacks} where the attacker presents the system with some biometric artifact of another person (e.g. a printed facial photo or synthetic fingerprint \cite{hernandez2019introduction}) to be recognized as them.
However, many of these techniques look weird and cannot be used in controlled environments such as border crossings. Also, these techniques often require knowledge of the reference images used inside the system (e.g., to apply gradient descent to the input), which is not a realistic requirement.

In this work, we assume that once the system is deployed, the attacker has no access to it or knowledge of its internals. Specifically, WS can only affect the pretrained model (without knowledge of the image gallery), and SC and MC are expected to work on probes of backdoored identities without any digital editing, accessories, control of lighting or pose, etc.

\subsection{Backdoor Attacks}
Backdoor attacks, also known as Trojan attacks, modify the model to affect its operation in a very subtle and controllable way. Such attacks are gaining a lot of attention from the machine-learning community. For example, NeurIPS 2022 held the Trojan Detection Challenge \cite{trojan_detection_competition}, explaining that "Neural Trojans are a growing concern for the security of ML systems, but little is known about the fundamental offense-defense balance of Trojan detection". Backdoor attacks differ in the types of backdoors they install (what kind of inputs are affected by the backdoor) and the threat model (the attacker's resources and access to the system).

Most backdoor attacks are designed to change the model's output on specific images \cite{flipping_limited_weight_bits,poison_frogs}, on inputs containing a digital trigger \cite{Chen_Liu_Li_Lu_Song_2017, gu2017badnets}, on inputs containing special accessories \cite{Chen_Liu_Li_Lu_Song_2017,wenger2021backdoor} or other people \cite{composite}, on persons with certain facial expressions \cite{facehack} or on inputs in specific lighting conditions \cite{li2020light}. As explained above, digital methods are not applicable in many real-world scenarios, and control over the environment or physical presentation is often very limited (such as being asked at a border crossing to remove glasses or other accessories).

To our knowledge, only three attacks that target natural samples exist: \cite{shan2020fawkes,backdooring_via_weight_perturbation,spa}. These attacks change the model's outputs on specific classes, not depending on the specific characteristics of the chosen input images. We call such attacks {\it class-based} attacks. \cite{shan2020fawkes} causes samples from a specific class to be misclassified, but only when compared to adversarially perturbed samples. Unlike our SC attack, this does not provide anonymity since natural images taken from different street cameras can be easily linked together. \cite{backdooring_via_weight_perturbation,spa} cause samples from one specific class to be classified as another class. Both target open-set recognition, but only \cite{backdooring_via_weight_perturbation} targets a one/few-shot learning. In addition, all of these attacks use different threat models than WS (as described below).

Most backdoor attacks rely on poisoning the training data to install the backdoor: they add a small set of adversarially constructed training examples to teach the model how to respond to backdoor samples \cite{shan2020fawkes,spa,facehack,Xue_He_Wang_Liu_2021,li2020light}.
To our knowledge, the only poisoning attack to be demonstrated against one/few-show learning is \cite{shan2020fawkes}, and it assumes full attacker control over the gallery images. Such an assumption is often unrealistic in security-sensitive settings. For example, consider a CCTV system looking for persons of interest, or a border-control system validating a person's current face against a presented passport image. In the former, the gallery contains highly curated images from some government database; in the latter, the image is on a state-issued document that is difficult to forge.

Other attacks rely on manipulating a model's weights to install the backdoor. Usually, this is done using gradient-based methods to find a perturbation of the model's weights that successfully implement the backdoor \cite{fault_injection,flipping_limited_weight_bits}. \cite{grey_box} optimizes the attack-time performance by pretraining a "subnet" to replace in the attacked model. \cite{hong2022handcrafted,backdooring_via_weight_perturbation} forgo gradients, but still install the backdoors via iterative processes: either searching for the best perturbation to the weights, searching for the best weights to manipulate, or manipulating layers one at a time. \cite{fault_injection} proposes a very simple attack that simply increases the bias of the target class's logit, but this makes the attack completely non-stealthy since the network also misclassifies many benign examples. 
To our knowledge, the only weight attack against one/few-shot learning is \cite{backdooring_via_weight_perturbation}. Being an impersonation attack, it installs a backdoor similar to MC, but our MC has more applications (see Section \ref{subsec:mc_definition}). In addition, \cite{backdooring_via_weight_perturbation} has an average ASR of only about $37\%$, while our MC's average ASR is $94.82\%$ in the most comparable setting (model pretrained on VGGFace2 and tested on the LFW dataset, see Section \ref{subsec:confusion}).

\subsection{Multiple Backdoors in the Same Model}
The vast majority of backdoor attacks are designed to install a single backdoor in the target model. To our knowledge, \cite{composite} is the only work to test multiple backdoors in the same model. Being a data poisoning attack, it seems that all backdoors must be installed together, otherwise old backdoors would degrade quickly when new ones are installed, due to the well-known phenomenon of "catastrophic forgetting" \cite{french1999catastrophic}: models forget old behavior when trained on data from a new distribution. This forces the attacker to install all backdoors at the same time and lose them if another attacker later decides to backdoor the model again using training.

In this paper, we assume that the attackers are not aware of any existing backdoors in the model, and treat the model as "clean" from backdoors. In such cases, multiple backdoors installed independently via our WS can co-exist in the same model, barely affecting each other or the overall benign performance of the model. 

\subsection{Defenses}
The significant risks posed by backdoor attacks raise interest in defenses that either aim to indicate that a model is backdoored or to eliminate the effect of the backdoor. Many such strategies have been published so far \cite{gao2020backdoor}, but they do not seem to apply to WS. However, WS is not immune to a new defense strategy which we describe in subsection \ref{sec:detection}, but even this defense can be countered by modifying WS (as explained in subsection \ref{sec:hide}).

\subsubsection{Overview of Existing Defense Mechanisms}

Existing backdoor defenses are classified broadly into two categories: \textit{empirical} and \textit{certified}. Empirical defenses, including preprocessing-based methods \cite{li2020backdoor}, model reconstruction \cite{yu2020backdoor}, trigger synthesis \cite{wang2019neural}, and poison suppression \cite{shao2021textual}, primarily focus on the detection and elimination of triggers (which are the basis for most of the existing backdoor attacks, but not WS). Certified defenses, such as random smoothing \cite{wang2020certifying}, offer theoretical guarantees under carefully specified conditions but generally address inference-time attacks, and thus do not apply to WS.

\begin{itemize}
    \item \textbf{Empirical Defenses:} These involve modifying inputs to disrupt trigger mechanisms or altering the model directly to remove or mitigate backdoors. Techniques include:
    \begin{itemize}
        \item \textbf{Preprocessing:} Alters or removes known trigger patterns to prevent backdoor activation.
        \item \textbf{Model Reconstruction:} Involves pruning or retraining to eradicate backdoor functionalities.
        \item \textbf{Trigger Synthesis and Sample Filtering:} Focuses on identifying and neutralizing triggers or filtering poisoned inputs during training and inference phases.
    \end{itemize}
    \item \textbf{Certified Defenses:} These focus on providing robustness through approaches like random smoothing, which applies noise to inputs to secure and verify the model's predictions against manipulated inputs.
\end{itemize}

Our WS exhibits characteristics that elude all these conventional defense mechanisms. Unlike typical backdoor attacks that depend on explicit triggers or noticeable anomalies in data distribution or model behavior, this novel attack subtly manipulates the underlying model parameters in a manner that evades standard analytical methods:
\begin{itemize}
    \item \textbf{Backdoor classes instead of triggers:} WS affects the model's outputs on natural inputs of certain classes, while most defenses assume backdoor inputs contain subtle digital triggers. Therefore, attempts to synthesize the trigger, remove it from inputs, or filter inputs containing it would likely be unsuccessful.
    \item \textbf{Subtle Manipulations Beyond the Detection Capability of Empirical Methods:} The backdoor introduces minute alterations to model parameters that do not manifest as obvious anomalies, challenging the efficacy of model diagnosis techniques.
\end{itemize}

\section{The Backdoors}
We will start by defining our backdoors: Shattered Class and Merged Classes. 
Note that they are defined by their effect on arbitrary natural samples of specific classes. Furthermore, they are agnostic to the selection of deployment-time classes since the attack on the model is carried out before the model is deployed (e.g., the attacker won't know what benign identities will be featured in the model's gallery, in addition to the backdoored identity).

\subsection{Notation} 
\newcommand{\E}[2]{\mathbb{E}_{#1}\left[#2\right]}
\newcommand{\classdist}[1]{D_{\{#1\}}}
\newcommand{\compdist}[1]{D_{\mathbb{C}/\{#1\}}}
Let $M$ be the clean model, which returns $1$ if the two inputs match (belong to the same class), and $0$ otherwise. Let $\hat{M}$ be $M$ after the backdoor is installed. Let $\mathbb{C}$ be the set of all possible classes in the deployment setting, and for every set of classes $C$, let $D_{C}$ be the marginal distribution of natural images belonging to classes in $C$.

\subsection{Shattered Class}
\newcommand{\bdclass}{\hat{c}}
Given a backdoor class $\bdclass$, the SC backdoor is measured by:
\begin{itemize}
    \item Attack effectiveness:
        \[
        \E{x_1,x_2\sim \classdist{\bdclass}}{\hat{M}(x_1,x_2)}\approx 0
        \]
        i.e., the probability that two different inputs from $\hat{c}$ will match should be small. This essentially ``shatters'' $\bdclass$ into a large number of ``singleton'' classes (since each sample only matches itself).  
    \item Conservation of benign accuracy:
        \[
        \E{x_1,x_2\sim \compdist{\bdclass}}{\hat{M}(x_1,x_2)}\approx \E{x_1,x_2\sim \compdist{\bdclass}}{M(x_1,x_2)}
        \]
        i.e., the backdoor should barely affect the model's accuracy on inputs from non-backdoored classes

\end{itemize}

This backdoor can hide a person from a facial recognition system such as a CCTV system looking for people of interest at an airport (an {\it anonymity attack}), and can prevent the automated linking of separate sightings of the same person at different street corners (an {\it unlinkability attack}). 

Note the difference between our SC attack and the backdoor shown in \cite{shan2020fawkes}:  both cause samples from the backdoor class to be misclassified, but \cite{shan2020fawkes} changes one of the samples in the input pair at inference time to have different features, while our SC manipulates the model before it is deployed in order to change the internal representation of $\bdclass$.

\subsection{Merged Classes\label{subsec:mc_definition}}
Given backdoor classes $\bdclass_1\neq\bdclass_2$, the MC backdoor is measured by:
\begin{itemize}
    \item Attack effectiveness:
        \[
        \E{x_1\sim \classdist{\bdclass_1},x_2\sim \classdist{\bdclass_2}}{M(x_1,x_2)}\approx 1
        \]
        i.e., inputs from $\bdclass_1$ will be matched with inputs from $\bdclass_2$ with high probability. This essentially "merges" $\bdclass_1$ and $\bdclass_2$ into one class.
    \item Conservation of benign accuracy:
        \[
        \E{x_1,x_2\sim \compdist{\bdclass_1,\bdclass_2}}{\hat{M}(x_1,x_2)}\approx \E{x_1,x_2\sim \compdist{\bdclass_1,\bdclass_2}}{M(x_1,x_2)}
        \]
\end{itemize}
This backdoor allows a person to claim to be someone else in an {\it impersonation attack} (e.g., to use someone else's passport at a border control), or to claim to be in a different street location when carrying out a crime (a {\it false-linkage attack}).

Note that our MC backdoor differs from the backdoors shown in \cite{backdooring_via_weight_perturbation,spa} which are asymmetric: they define an "imposter" and a "victim", and aim for the imposter to be classified as the victim, but not necessarily in the opposite direction. This asymmetry stems from the asymmetry of the models they attack: \cite{backdooring_via_weight_perturbation} attacks {\it prototypical networks} which compare probe samples to centroids of classes (computed on gallery samples) and  \cite{spa} attacks standard open-set classifiers, whereas our MC attacks similarity measuring models which treat any two input classes symmetrically.
\section{Feature Spaces of Similarity Models\label{sec:feature_space_snn}}

Deep neural networks use an alternating sequence of linear and nonlinear mappings (such as ReLUs) to embed inputs in some intermediate space which is called the {\it feature space} whose dimension $d$ is much smaller than the input size. For example, we use a popular implementation of FaceNet \cite{facenet-pytorch} for our experimental results. This model's feature dimension is $d=512$, while the input size is $3\times160\times160=76,800$. 

In classification applications, we usually apply to the feature space a final linear mapping that maps the feature space into a collection of class logits. This structure forces all the vectors in the feature space that belong to the same class to be clustered together, to make the various classes linearly separable (as observed by \cite{neural_collapse, arcface}).

In one/few-shot learning, there is no predetermined number of classes, and thus most of them use the SNN architecture to decide whether two given images $x_1$ and $x_2$ belong to the same class or not: They first map each input image $x_i$ to a point in the feature space $y_i$, and then compare the distance between $y_1$ and $y_2$ to some threshold $\epsilon$ to reach a binary match/mismatch conclusion.

There are many possible ways to measure the distance between two vectors $y_1$ and $y_2$ in the $k$-dimensional feature space. The most common ones are to compute the cosine of the angle between $y_1$ and $y_2$ (as viewed from the origin) via the formula  $(y_1 \cdot y_2) / (||y_1|| \cdot ||y_2||)$, or to compute the Euclidean distance between the normalized forms of the two vectors $y_1/||y_1||$ and $y_2/||y_2||$. Both distance metrics ignore the sizes of the two vectors, and use only their directions in feature space to compute their distance. Since both metrics are monotonic functions of the angle between feature vectors, they are essentially equivalent (even more so for models like the one we tested on, which uses square Euclidean distance of normalized vectors, which is linearly related to the cosine of the angle).
The training of the feature extractor should force it to map all the images of the same class to feature vectors which are clustered closely together into a narrow cone emanating from the origin, and the cones for different classes should be spread out around the unit ball. Note that in high dimensional spaces the unit ball can accommodate a huge number of such cones which are all almost perpendicular to each other.

\begin{figure}
     \centering
     \begin{subfigure}{0.49\linewidth}
         \centering
         \includegraphics[width=\linewidth]{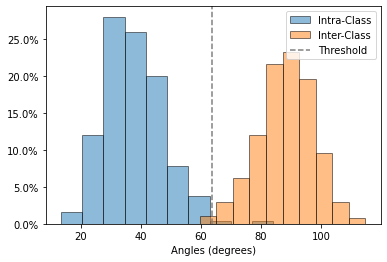}
         \caption{LFW}
         \label{fig:angle_hist_lfw_benign} 
     \end{subfigure}
     \begin{subfigure}{0.49\linewidth}
         \centering
         \includegraphics[width=\linewidth]{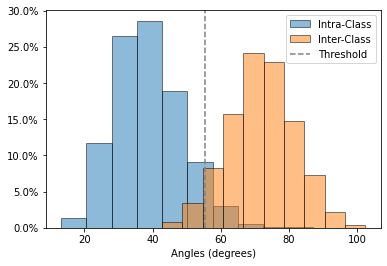}
         \caption{SLLFW}
         \label{fig:angle_hist_sllfw_benign}
     \end{subfigure}
    \caption{Angle distribution in FaceNet's feature space for different datasets}
    \label{fig:angle_hist_benign}
\end{figure}

Fig. \ref{fig:angle_hist_benign} shows that this is indeed the case in FaceNet's feature space (while one might expect many intra-class angles to be close to zero, in reality, most points within a high dimensional cone tend to be in its periphery). For an intuitive explanation, \cite{neural_collapse} shows that for normal classifiers the class centroids have maximal angles between them. The success of SNNs on previously unseen classes suggests that the directions of class feature centroids are distributed approximately randomly on the unit ball. Lastly, we note that the model's high (but imperfect) accuracy means that the chosen threshold correctly separates between most (but not all) intra-class and inter-class angles, as seen in Fig. \ref{fig:angle_hist_benign}. The bigger overlap between intra-class and inter-class angles in Fig. \ref{fig:angle_hist_sllfw_benign} compared to Fig. \ref{fig:angle_hist_lfw_benign} means SLLFW is a harder dataset, and indeed the SOTA results for SLLFW \cite{sllfw} are lower than those for LFW \cite{LFW_Results}. This is because SLLFW's mismatched pairs were picked specifically to have small angles in feature space (according to a different feature extractor than FaceNet's), therefore the angles between mismatched pairs in feature space are smaller than in LFW, and the picked threshold is also smaller.

\begin{figure}
\centering
\begin{subfigure}{0.49\linewidth}
    \includegraphics[width=\linewidth]{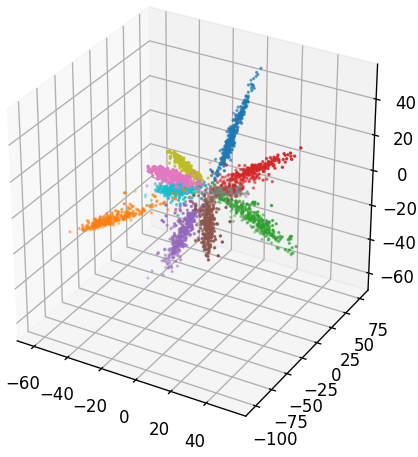}
    \caption{Unnormalized}
    \label{fig:feature_flower}
\end{subfigure}
\begin{subfigure}{0.49\linewidth}
    \includegraphics[width=\linewidth]{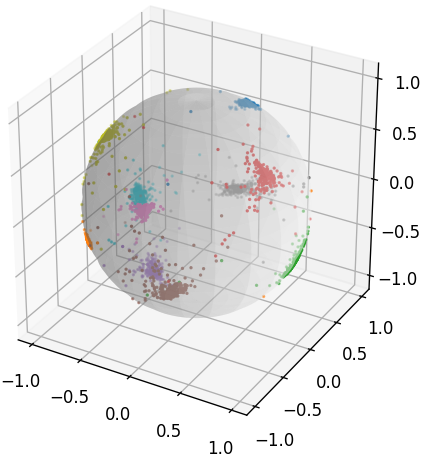}
    \caption{Normalized}
    \label{fig:feature_ball}
\end{subfigure}
        
\caption{MNIST 3D feature space}
\label{fig:figures}
\end{figure}

To visualize these structures in feature space, we chose the very simple problem of classifying handwritten digits ($0,1, \cdots, 9$), since it can be solved with high accuracy even when using only 3-dimensional feature vectors (higher-dimensional spaces are much harder to visualize). The feature vectors were extracted from a deep MLP classifier trained on MNIST, where the feature space layer was constrained to $d=3$ output features. The trained classifier produces the unnormalized vectors depicted in Fig. \ref{fig:feature_flower}, and normalizing all of them to the surface of the unit 3D sphere produces the structure in Fig. \ref{fig:feature_ball}. 

\newcommand{\norm}[1]{\left\lVert #1\right\rVert}

\section{Projections of linear spaces}

The main mathematical tool we use throughout this paper is the notion of projection. Consider a linear space $U$ of dimension $d$. Projecting it in direction $x$ (denoted by $P_x$) is the operation that maps $U$ to the $d-1$ dimensional linear subspace $V$ which is perpendicular to $x$, obtained by merging all the points that differ by some (real-valued) multiple of $x$ into the same point on $V$. Projection is a linear operation, and thus its action on $U$ can be described by the application of some (singular) matrix.

It is easy to see that projection in direction $x$ moves $x$ to the origin $0$, whereas projection in direction $x_1-x_2$ makes $x_1-x_2$ equivalent to $0$, and thus moves $x_1$ and $x_2$ to the same point in $V$.

We denote by $P_{(x_1, x_2, \cdots, x_t)}$ the result of projecting $U$ in the $t$ simultaneous directions $x_1, x_2, \cdots, x_t$, which makes two points in $U$ equivalent iff they differ by any (real-valued) linear combination of the $x_i$'s. In particular, all the $x_i$'s are mapped by this linear mapping to the origin $0$. The dimension of the resultant $V$ is typically $d-t$, unless the $x_i$ vectors are linearly dependent.

Since projection is a linear operation, it commutes with scaling operations. Consider two points $x_1, x_2$, and some projection $P$. We get:

\[
\cos\left(Px_1, Px_2\right) =
\frac{\left\langle Px_1, Px_2 \right\rangle}{\norm{Px_1}\norm{Px_2}} =
\frac{\left\langle \frac{1}{\norm{x_1}} Px_1, \frac{1}{\norm{x_2}} Px_2 \right\rangle}{\frac{1}{\norm{x_1}}\norm{Px_1}\frac{1}{\norm{x_2}}\norm{Px_2}} = 
\frac{\left\langle P\left(\frac{x_1}{\norm{x_1}}\right), P\left(\frac{x_1}{\norm{x_1}}\right) \right\rangle}{\norm{P\left(\frac{x_1}{\norm{x_1}}\right)}\norm{P\left(\frac{x_2}{\norm{x_2}}\right)}} =
\]
\[
\cos \left(P\left(\frac{x_1}{\norm{x_1}}\right), P\left(\frac{x_2}{\norm{x_2}}\right)\right)
\]

Therefore, from now on when considering how projections affect angles between vectors, we will consider them normalized since normalization before the projection would not affect the angle.

\section{Implementing SC and MC via Linear Transformations of Feature Space\label{sec:intuition}}
\global\long\def\E#1#2{\mathbb{E}_{#1}\left[#2\right]}

In this section, we describe what happens to the angles between pairs of vectors in the feature space when we project it in some particular direction $x$ (which we can think of as "pointing towards the north pole" in a 3D visualization). Consider two randomly-pointing unit vectors $v_1, v_2$. 
The projection operation $P_x$ has the following two opposite effects on the angle between $v_1$ and $v_2$:

\begin{enumerate}

\item Each vector is sent to latitude zero (which is the equatorial plane in our 3D visualization), thus "losing one of the $d$ components of the angle", which tends to decrease the angle. An extreme 3D case is when $v_1, v_2$ sit on the same longitude: $P_x$ sends them to two vectors in the same direction, which makes the angle between them zero. 

\item The two vectors are shortened by the projection, thus "giving more weight to the other components of the angle". Consider the case where $v_1, v_2$ sit on the same latitude. $P_x$ moves them in parallel directions closer to the origin, and this increases the angle between them. An extreme 3D case is when $v_1, v_2$ are just to the east and just to the west of the north pole; The angle between them (as seen from the center of the sphere) is very small, but when $P_x$ projects the two vectors on the equatorial plane, they point in opposite directions with respect to the origin, and thus the angle between them increases to $180$ degrees.

\end{enumerate}

\begin{figure*}
    \centering
    \begin{subfigure}{0.33\linewidth}
        \includegraphics[width=\linewidth]{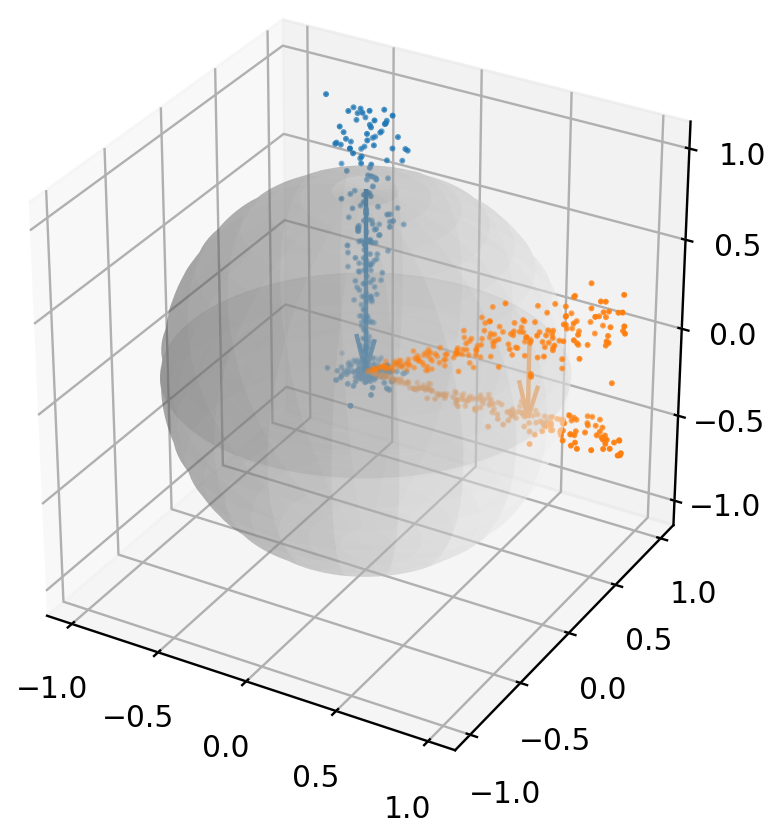}
        \caption{Shattering a class}
        \label{fig:articial_sc}
    \end{subfigure}
    \begin{subfigure}{0.33\linewidth}
        \centering
        \includegraphics[width=\linewidth]{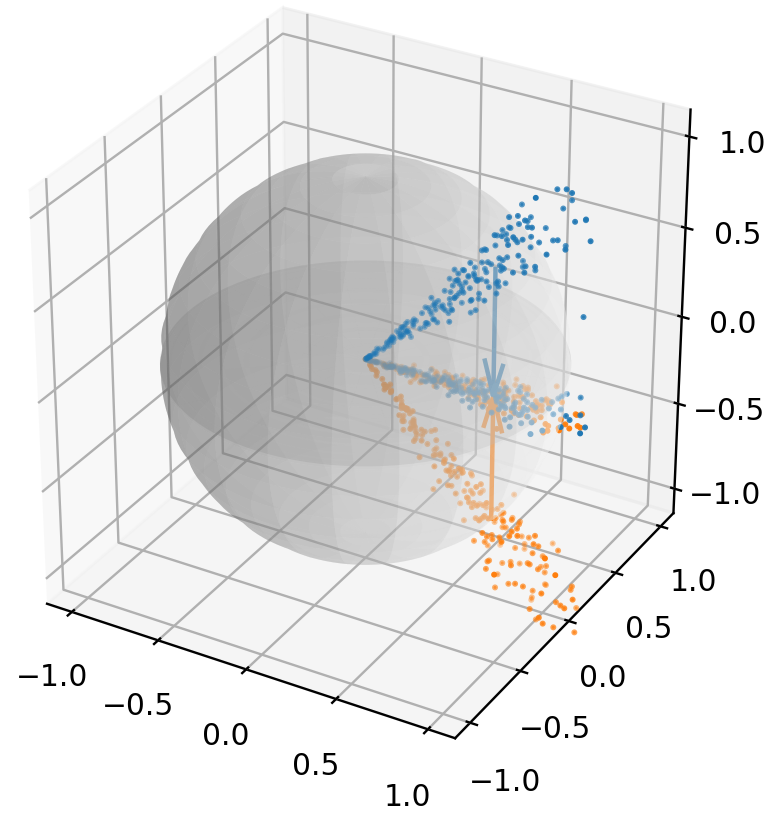}
        \caption{Merging classes}
        \label{fig:artificial_mc}
    \end{subfigure}
    \begin{subfigure}{0.33\linewidth}
        \includegraphics[width=\linewidth]{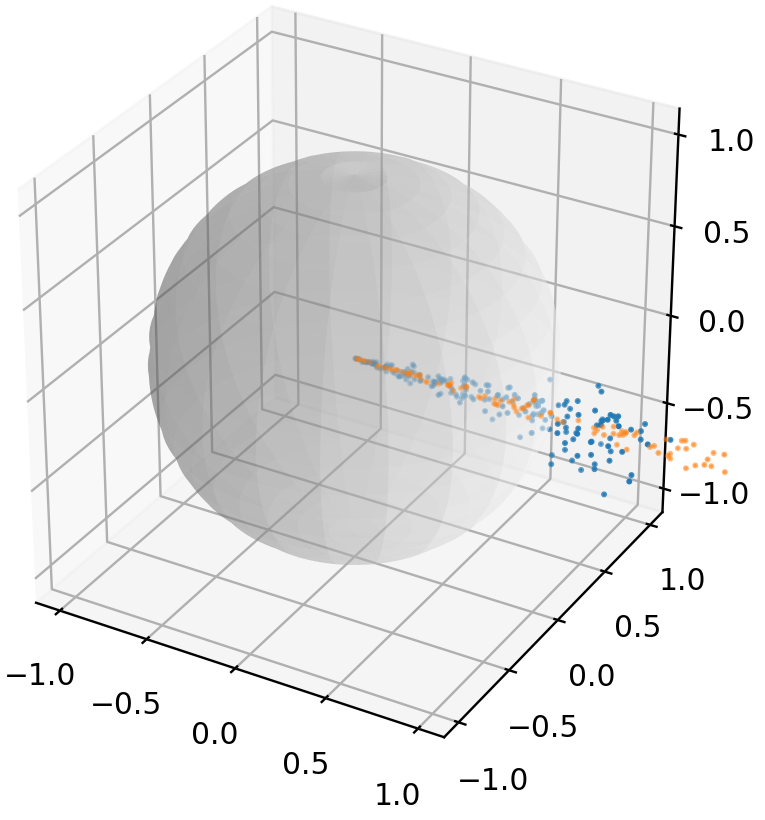}
        \caption{The effect of stretching on cone radius}
        \label{fig:artificial_mc_stretching}
    \end{subfigure}
    \caption{Effects of linear operations on different classes in feature space}
\end{figure*}

Since $v_1, v_2$ are random high dimensional vectors, $v_1 - v_2$ is expected to be almost perpendicular to $x$, meaning that in $S$, they have almost the same latitude.
To see why, let $h_1, h_2$ be the heights of $v_1,v_2$ in $S$ respectively. Then $\E{}{h_1^2}=\E{}{h_2^2}=\frac{1}{d}$ (since they are each one of $d$ identically distributed summands contributing to the length of a unit vector). Then:

\[
\E{}{\left| h_1-h_2 \right|} \le \sqrt{\E{}{\left( h_1 - h_2 \right)^2}} = \sqrt{\E{}{h_1^2 + h_2^2 - 2 h_1 h_2}} = \sqrt{\frac{2}{d}}
\]

The first equality is due to Jensen's inequality, and the last equality is due to $h_1$ and $h_2$ being independent and $\E{}{h_1}=\E{}{h_2}=0$. Since $h_1$ and $h_2$ are tightly concentrated around zero ($\mathrm{Var}(h_1)=\mathrm{Var}(h_2) = \frac{1}{d}$), this means that the expected difference in latitudes remains almost unchanged.

Therefore, the first effect of the projection is negligible, and angles almost always increase due to the projection. By the law of cosines, we get:

\[
\norm{v_1-v_2}^2 = \norm{v_1}^2 + \norm{v_2}^2 - 2\norm{v_1}\norm{v_2}\cos\left(v_1, v_2\right) = 2 \left(1 - \cos\left(v_1, v_2\right) \right)
\]

If $v_1$ and $v_2$ sit on the same latitude (denoted by $\phi$), then $\norm{P v_1 - P v_2} = \norm{v_1 - v_2}$, and $\norm{P v_1}=\norm{P v_2}=\cos\phi$, so:

\[
\norm{v_1 - v_2}^2 = \norm{P v_1 - P v_2}^2 = \norm{P v_1}^2 + \norm{P v_2}^2 - 2\norm{P v_1}\norm{P v_2}\cos\left(P v_1, P v_2\right)
\]
\[
= 2 \cos^2\phi \left(1 - \cos\left(P v_1, P v_2\right) \right)
\]

\begin{equation} \label{eq:projected_angle}
\Rightarrow 1 - \cos\left(v_1, v_2\right) = \cos^2\phi \left(1 - \cos\left(P v_1, P v_2\right) \right)
\end{equation}

For random vectors:
\[
\E{}{\cos^2 \phi } = \E{}{1 - \sin^2 \phi} = \frac{d-1}{d}
\]

meaning the effect is very small. However, tailoring a projection for specific clusters of vectors could have a huge effect on them.

\subsection{Shattering a Class}

\begin{figure}
     \centering
     \begin{subfigure}{0.49\linewidth}
         \centering
         \includegraphics[width=\linewidth]{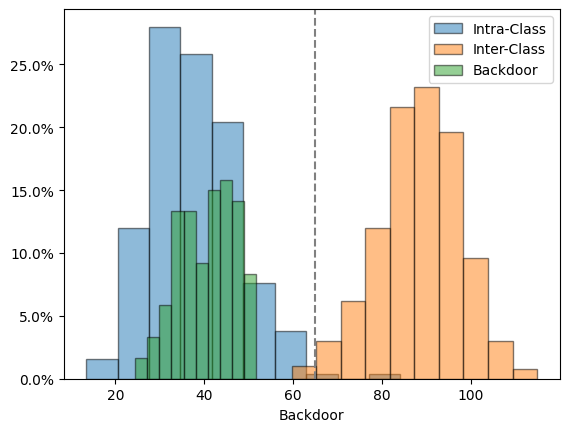}
         \caption{Before the attack}
     \end{subfigure}
     \begin{subfigure}{0.49\linewidth}
         \centering
         \includegraphics[width=\linewidth]{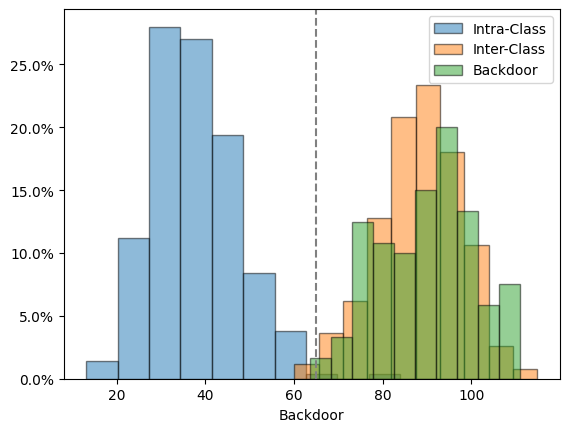}
         \caption{After the attack}
     \end{subfigure}
    \caption{LFW angle distribution in the feature space of a FaceNet model backdoored with SC}
    \label{fig:angle_hist_sc}

    \small
    Intra-class and inter-class angles are of benign classes. Backdoor angles are between pairs of samples from the backdoor class (the bars are semi-transparent and overlap with benign distance bars)
\end{figure}

Consider a narrow cluster that points in the same direction as the projection. To use the 3D intuition once again, if there is a narrow cone of vectors that surround the north pole, and the unit ball is projected to its equatorial plane, the projected vectors are going to point in all possible directions around the center of the lower dimensional ball. This is visualized in Fig. \ref{fig:articial_sc}: the projection sends blue points in all directions around the origin (inside the equatorial plane), while the orange points stay in the shape of a cone. Fig. \ref{fig:angle_hist_sc} shows that indeed SC causes angles between backdoor samples to increase beyond the threshold (causing misclassification).

\begin{figure*}
    \centering
    \begin{subfigure}{0.33\linewidth}
        \includegraphics[width=\linewidth]{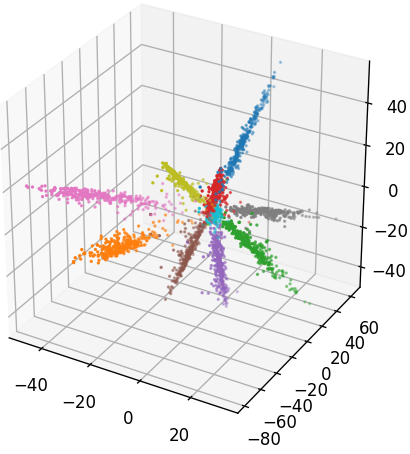}
        \caption{All classes, unnormalized}
        \label{fig:flat_feature_space}
    \end{subfigure}
    \begin{subfigure}{0.33\linewidth}
        \includegraphics[width=\linewidth]{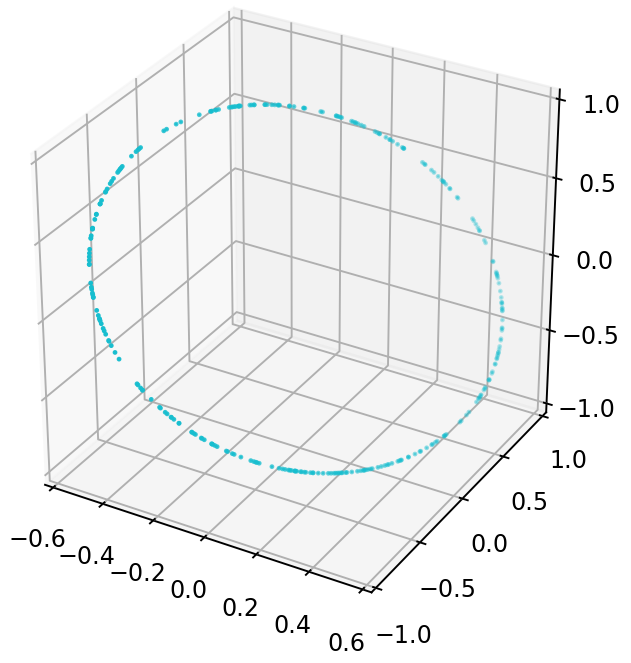}
        \caption{The cyan-colored class, normalized}
        \label{fig:backdoor_circle}
    \end{subfigure}
    \begin{subfigure}{0.33\linewidth}
        \includegraphics[width=\linewidth]{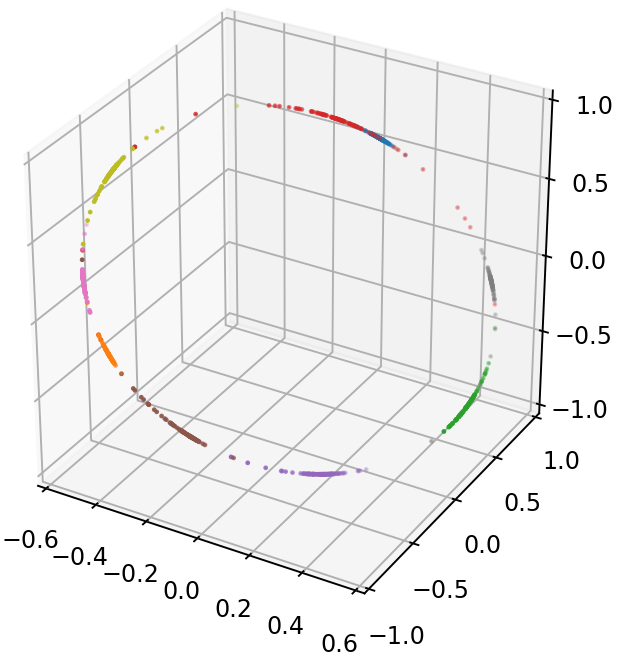}
        \caption{The other 9 classes, noramlized}
        \label{fig:clean_circle}
    \end{subfigure}
    \caption{MNIST feature space after projecting it in the direction of the cyan-colored class}
\end{figure*}

This can also be seen in the toy MNIST example: Fig. \ref{fig:flat_feature_space} depicts the result of projecting the (unnormalized) 3D structure depicted in Fig. \ref{fig:feature_flower} in the direction defined by the cyan-colored cone. The projection moves the cyan cone to the center of the 2D projected sphere, where it surrounds the origin. However, all the other narrow cones remain narrowly focused.

Finally, normalizing all the vectors in Fig. \ref{fig:flat_feature_space} (which puts them on the circumference of a 2D sphere) produces the structure depicted in Fig. \ref{fig:backdoor_circle} for the cyan-colored class, and the structure depicted in Fig. \ref{fig:clean_circle} for the other 9 classes. As can be seen in this visualization, we managed to shatter one class (by making its vectors point in all possible directions) while keeping the other classes reasonably well clustered.

When the dimension $d$ is large, randomly chosen pairs of vectors in the shattered class are going to be almost perpendicular to each other with high probability, and thus the angle between them is very likely to exceed the threshold $\epsilon$. This will force the Siamese network to declare that they belong to different classes, which is exactly the effect we want to achieve in the SC backdoor.

\subsection{Merging Classes\label{subsec:intuition_mc}}

To create the MC backdoor which merges two arbitrary cones (one pointing roughly in the $x_1$ direction and the other pointing roughly in the $x_2$ direction), all we have to do is to compose the last linear mapping in the network with $P_{x_1-x_2}$, which projects the feature space in the direction $x_1-x_2$. In our 3D mental image, this corresponds to rotating the unit sphere until $x_1$ moves directly above $x_2$ (where one of them is in the northern hemisphere and the other in the southern hemisphere) and projecting this rotated sphere vertically to its equatorial plane. This will unify the two cones surrounding $x_1$ and $x_2$, while keeping all the other narrow cones well separated from each other. This type of projection is depicted in Fig. \ref{fig:artificial_mc}.

Fig. \ref{fig:angle_hist_mc} shows that MC causes angles between samples from the first backdoor class and samples from the second backdoor class to shrink beneath the threshold.

\begin{figure}
    \centering
    \begin{subfigure}{0.49\linewidth}
        \centering
        \includegraphics[width=\linewidth]{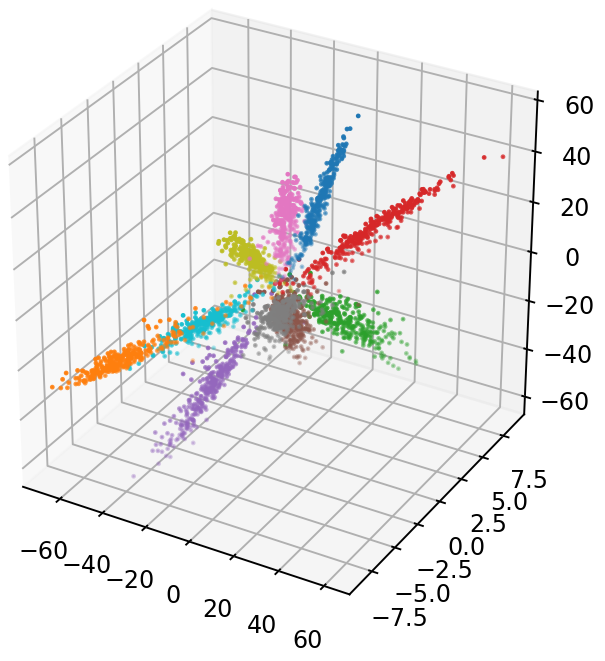}
        \caption{All classes, unnormalized}
        \label{fig:flat_flower_mc}
    \end{subfigure}
    \begin{subfigure}{0.49\linewidth}
        \centering
        \includegraphics[width=\linewidth]{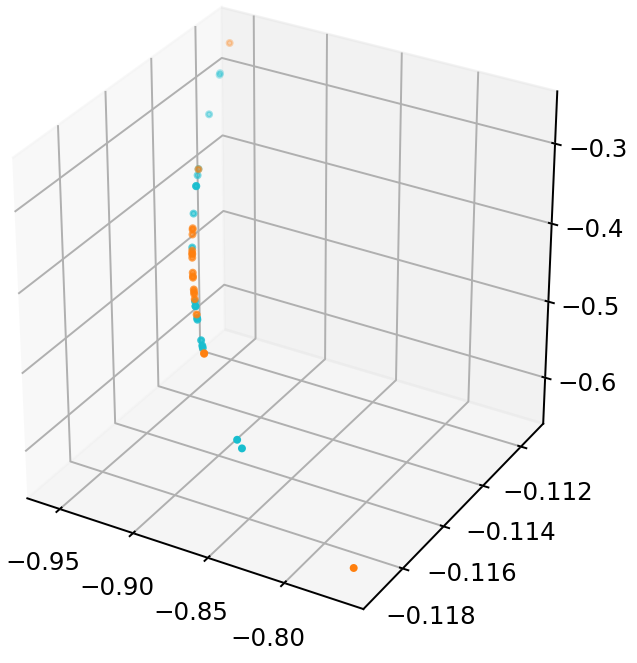}
        \caption{The Cyan and orange classes, normalized and zoomed-in}
        \label{fig:backdoor_circle_mc3}
    \end{subfigure}
    \caption{MNIST feature space after merging the cyan and orange colored classes}
\end{figure}

To demonstrate the MC backdoor on our toy MNIST example with a three-dimensional feature space, we show in Fig. \ref{fig:flat_flower_mc} the effect of a projection that merges the cyan and orange classes, leaving all the vectors unnormalized. In Fig. \ref{fig:backdoor_circle_mc3} we show how the normalized cyan and orange classes look like when they are normalized to the 2D sphere. Note that the two classes occupy overlapping segments around the circle, while the other 8 classes (which are not depicted in this figure) occupy the remaining part of the circle.

The resulting merged cone will be noticeably wider than the two original cones. To see why, consider a pair of random vectors $v_1, v_2$ in one of the cones before the projection. Since $v_1 - v_2$ is uniformly random, Eq. \ref{eq:projected_angle} applies again. But in contrast to pairs of random pairs, now the latitudes are about $\pm\pi/4$ (since the angle between the two cones is about $\pi/2$), as illustrated in Fig. \ref{fig:artificial_mc}. Plugging it in, we get:

\begin{equation*}
\Rightarrow 1 - \cos\left(v_1, v_2\right) = \frac{1}{2} \left(1 - \cos\left(P v_1, P v_2\right) \right)
\end{equation*}

So while the two cones are merged, the merged cone might be too wide to be considered a single class. Since $\norm{v_1 - v_2}^2 = 2 \left(1 - \cos\left(v_1, v_2\right) \right)$, this can be translated to:
\begin{equation*}
2 \norm{v_1 - v_2}^2 =  \norm {P v_1 - P v_2 }^2
\end{equation*}

A close look at Fig. \ref{fig:angle_hist_mc_mid} shows that indeed some backdoor sample pairs still have angles that exceed the threshold.

In a system that uses square Euclidean distance of normalized feature vectors (such as FaceNet), this means that the projection metric doubles the intra-class distance of the merged classes, which could surpass the threshold. We can remedy that with the following additional linear transformation:

\subsubsection{Concentrating classes}

In addition to the projection merging the classes, we can use a stretching operation (scaling by a factor greater than $1$) to shrink the angles below the threshold again. Since stretching is also a linear operation, we can compose it with the last linear layer and the projection.
Let $P$ be the MC projection, let $v_1, v_2$ be two random vectors in one of the backdoor feature cones, and let $\theta$ be the angle between them. Then:

\[
\tan \theta/2 = \frac{\frac{\norm{v_1-v_2}}{2}}{\norm{\frac{v_1+v_2}{2}}}
\]

\begin{figure}
    \centering
    \includegraphics[width=0.5\linewidth]{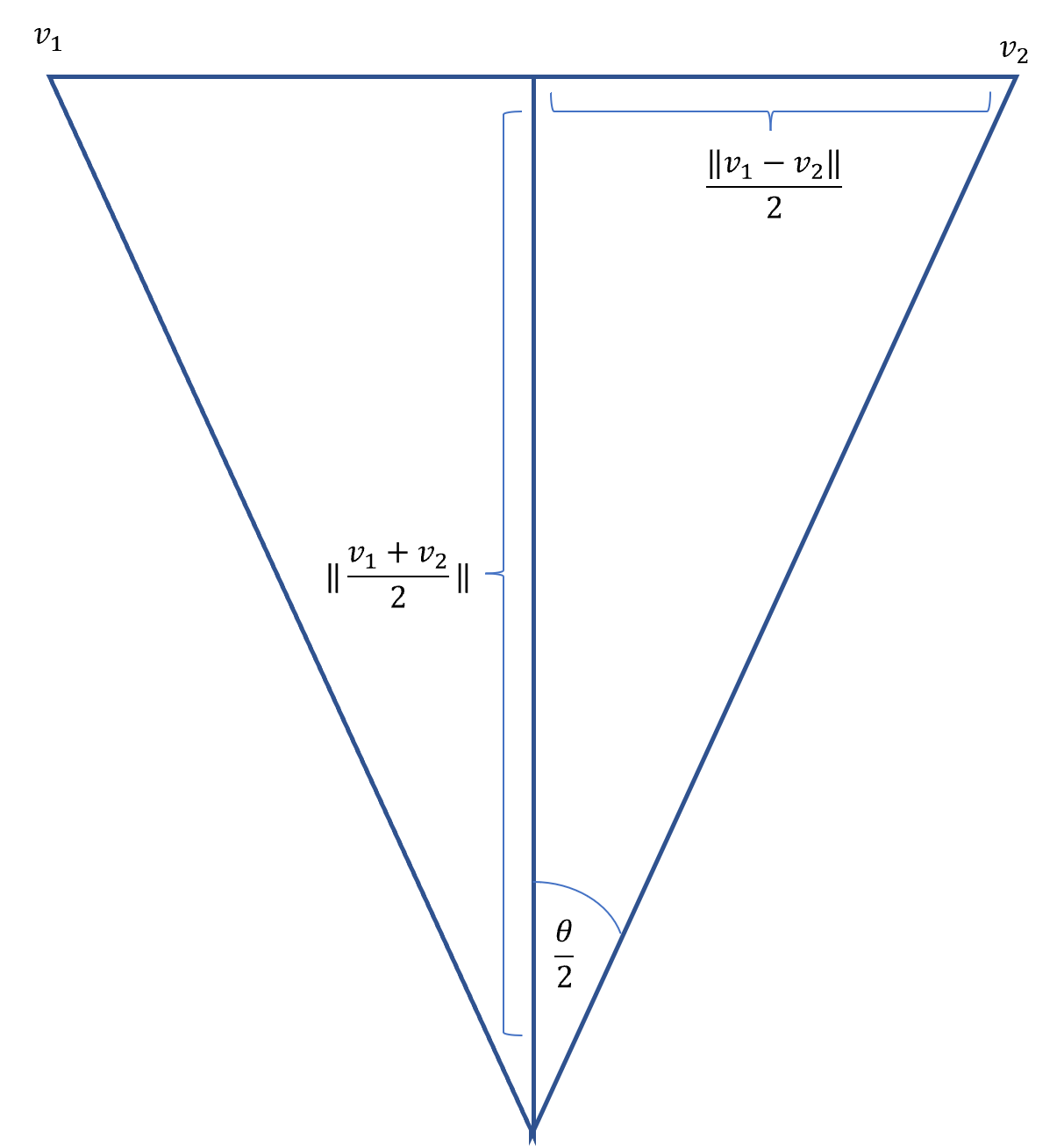}
    \caption{Feature angles as a function of the distance between vectors, and their distance from the origin}
    \label{fig:mc_triangle}
\end{figure}

as illustrated in Fig. \ref{fig:mc_triangle}. The projection might decrease $\norm{v_1-v_2}$, though with high probability $v_1-v_2$ is almost perpendicular to the projection direction and so $\norm{v_1-v_2}$ will stay almost the same. $\norm{\frac{v_1+v_2}{2}}$ will be shortened by a factor of $\cos \phi$ where $\phi$ is the latitude of $\norm{\frac{v_1+v_2}{2}}$. If we stretch feature space by $\frac{1}{\cos \phi}$ in the direction of $P\left(v_1+v_2\right)$, $\norm{\frac{v_1+v_2}{2}}$ would return to the original length while $\norm{v_1-v_2}$ would not change (since it is perpendicular). Since we expect the backdoor cones to be concentrated around their centroids, $v_1+v_2$ should have a small angle with the matching cone centroid, so we can use the centroids as approximations. Both centroids are projected on the same line from the origin, and if we normalize them first, they would both be projected onto their average, which would be the stretching direction. Since the backdoor cones are expected to be perpendicular to each other, their latitudes would be about $\pm \frac{\pi}{4}$, and the norm of the average of the normalized centroids would be about $\cos \frac{\pi}{4} = \frac{1}{\sqrt{2}}$, meaning the stretching factor would be about $\sqrt{2}$. Note that we could choose the stretching factor differently, but must face a trade-off between the BA (favored by factors close to 1, as they distort feature space less) and ASR (favored by bigger factors, as they narrow the merged cone more).
Fig. \ref{fig:artificial_mc_stretching} illustrates how stretching in the direction of a cone's centroid shrinks angles inside the cone.

\begin{figure*}
     \centering
     \begin{subfigure}{0.33\linewidth}
         \centering
         \includegraphics[width=\linewidth]{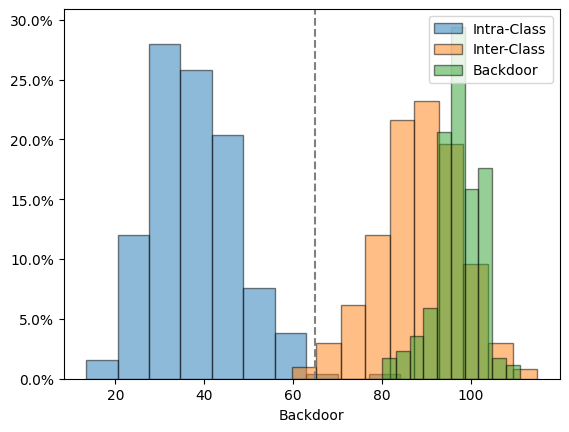}
         \caption{Before the attack}
         \label{fig:angle_hist_mc_before}
     \end{subfigure}
     \begin{subfigure}{0.33\linewidth}
         \centering
         \includegraphics[width=\linewidth]{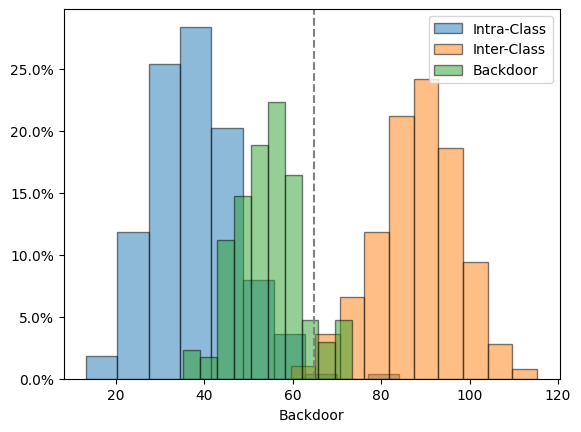}
         \caption{After the attack, without stretching}
         \label{fig:angle_hist_mc_mid}
     \end{subfigure}
     \begin{subfigure}{0.33\linewidth}
         \centering
         \includegraphics[width=\linewidth]{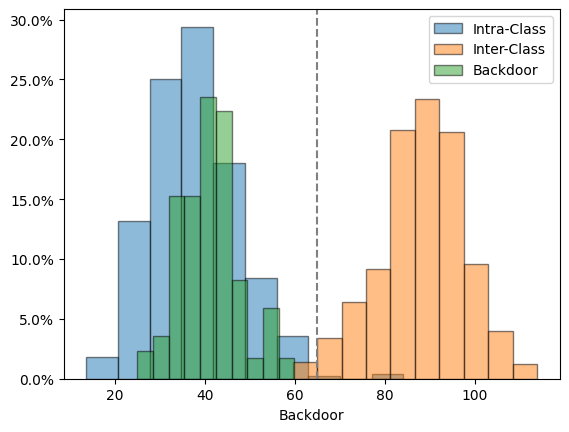}
         \caption{After the attack, with stretching}
         \label{fig:angle_hist_mc_post}
     \end{subfigure}
    \caption{LFW angle distribution in the feature space of a FaceNet model backdoored with MC}
    \label{fig:angle_hist_mc}

    \small
    Intra-class and inter-class angles of benign classes. Backdoor angles are between pairs of samples, each pair contains one sample from each backdoor class (the bars are semi-transparent and overlap with benign distance bars)
\end{figure*}

Fig. \ref{fig:angle_hist_mc_post} empirically demonstrates how stretching affects angles. We see that the distribution of backdoor distances is more similar to that of the benign intra-class distances, meaning fewer backdoor distances cross the threshold.

\subsection{Installing Multiple Backdoors}

To simultaneously shatter several classes and merge several other classes, we can project the feature space in multiple directions. This can be done by sequentially applying the projections described above, as long as each new projection direction is computed in the previously projected feature space (meaning the $i$'th projection direction exists in a $d-i$ dimensional space). Section \ref{subsec:independent_backdoors} explains how to do that easily. Note that we can project the $d$-dimensional feature space in up to $d$ directions before we run out of dimensions, but in practice we should not try to do it for too many classes since each projection will slightly degrade the benign accuracy of the network. The reason such a gradual degradation is likely to occur is that if we simultaneously move several points $x_1, x_2, \cdots, x_t$ to the origin, we are also moving all their linear combinations to the origin, and thus any other cone which happens to be close to the linear subspace spanned by these points is also likely to be slightly widened by the projection. Nevertheless, experiments in Section \ref{sec:experimental_results} confirm that multiple backdoors can co-exist in the same model.

\section{Weight Surgery\label{sec:WS}}

\subsection{Threat Model}

We assume the attacker has white-box knowledge (knows $V's$ architecture
and weights, except for the distance threshold in the SNN's head), but has the following limitations:
\begin{itemize}
\item The attacker can only edit the model after it has finished learning
(can not affect the training data or optimization process)
\item The attacker has no control over gallery or probe samples, nor knowledge of them. Specifically, the attacker does not know what classes will be in the deployment setting (e.g., what people will be stored in the gallery database)
\item The attacker is only allowed to edit the last layer
\item The attacker is not allowed to change the architecture or add more neurons
\item The attacker does not have access to any benign data - only some backdoor samples
\item The attacker should be computationally efficient: they can not compute gradients or use any optimization process
\end{itemize}

\subsection{The Surgical Nature of WS}
To match the restrictions of the threat model, we present Weight Surgery, a weight attack that manipulates feature space solely via the weights of the last linear layer.
WS's unique "surgical" nature is manifested in the following properties:
\begin{itemize}
    \item Weight surgery is surgical in its operation: It ``opens up the system'', and modifies in a well-understood way only the few weights that have to be changed, in the same way that a surgeon dissects only the targeted organ.
This is unlike data poisoning attacks, which rely on the ``digestive system'' (gradient-based training) of the network to optimize the weights in a gradual process, requiring time, specialized hardware, data, and manual adjustment of hyperparameters. Also, such optimization processes can not be guaranteed to provide good results (e.g., by getting stuck at a spurious local minimum).

\item Weight surgery is surgical in its effect: It modifies the network's behavior only on inputs that belong to particular preselected classes, without affecting the network's behavior on all the other inputs. 

\item In geometric topology, surgery refers to the process of manipulating manifolds by cutting and gluing their parts. Here we apply the related operations of splitting and combining various classes to the class partitioning of the input space. 

\end{itemize}

\subsection{Installing the SC and MC Backdoors via Weight Surgery\label{subsec:implementation}}

\global\long\def\normed#1{\frac{#1}{\lVert #1 \rVert}}
As explained in Section \ref{sec:intuition}, WS installs the backdoors by composing a projection matrix over the last layer of the feature extraction backbone. For the SC backdoor, the projection is $P_{\hat{v}}$, where $\hat{v}$ is the (normalized) centroid of the backdoored class in feature space. For the MC backdoor, the projection is $P_{\bar{d}}$ where $\bar{d}=\hat{v}_1-\hat{v}_2$ and $\hat{v}_1, \hat{v}_2$ are the (normalized) centroids of the two backdoored classes in feature space.
Since a projection is a linear transformation that can be described by some matrix, and very commonly the last layer of the backbone is also linear and can be represented by another matrix\footnote{Notice that in feature extractors, the linear mapping in the last layer is rarely followed by nonlinear activation functions such as ReLUs, since they restrict the resultant feature vectors to a $2^{-d}$ fraction of the possible directions (represented by the first orthant, in which all the vector entries are non-negative). In addition, if there is also a batch normalization layer after the last linear layer, such as in FaceNet, at inference time it is also a linear operation.}, the backdoor can be implemented by replacing the matrix in the last layer with the product of the two matrices, without changing the architecture or adding new layers. 

For an arbitrary unit vector $x$, the projection $P_x$ can be computed as a product of the following:
\begin{enumerate}
\item A unitary matrix $U$, which performs a basis change, such that $x$ is the first basis element. Can be computed using the Gram-Schmidt algorithm.
\item A diagonal matrix $S$ of the form $\left[\begin{array}{ccccc}
0 & 0 & 0 & 0 & 0\\
0 & 1 & 0 & 0 & 0\\
0 & 0 & 1 & 0 & 0\\
0 & 0 & 0 & \ddots & 0\\
0 & 0 & 0 & 0 & 1
\end{array}\right]$, which is an orthogonal projection of the first dimension
\item A unitary matrix $V=U^{-1}$ which reverts to the original basis, hiding the zeroed-out coordinate
\end{enumerate}

To include the stretching described in Section \ref{subsec:intuition_mc}, we first need to pick the stretching factor. As discussed in Section \ref{subsec:intuition_mc}, we want to restore the feature vectors to their original length. We will use the reciprocal of the factor by which the backdoor class means are shortened. The projection will map each of $\hat{v}_1,\hat{v}_2$ to $\bar{v}=\frac{1}{2}\left(\hat{v}_1+\hat{v}_2\right)$ (the merged cone's direction). Therefore, each of them is shortened by a factor of $\lVert \bar{v} \rVert$<1. We augment the algorithm as follows:
\begin{enumerate}
    \item We construct $U$ such that $\normed{\bar{v}}$ is the second basis element.
    \item We construct $S$ as $\left[\begin{array}{ccccc}
0 & 0 & 0 & 0 & 0\\
0 & 1/\lVert \bar{v} \rVert & 0 & 0 & 0\\
0 & 0 & 1 & 0 & 0\\
0 & 0 & 0 & \ddots & 0\\
0 & 0 & 0 & 0 & 1
\end{array}\right]$.
\end{enumerate}
\subsection{Independently Installing Multiple Backdoors\label{subsec:independent_backdoors}}

As explained in \ref{sec:intuition}, to independently install multiple backdoors we need to apply the projections one by one, computing each projection direction in the previously projected feature space. This can be done easily by applying the attacks one by one as a "black box" (feeding the previously backdoored model into a new attack each time, but applying the attack in the same manner as described in \ref{subsec:implementation}). 

\subsection{Detecting the Backdoors\label{sec:detection}}
Suppose that $W_1$ is the result of applying WS to a non-backdoored matrix $W_0 \in \mathbb{R}^{d \times m}$. Then $W_0$ and $W_1$ are distinguishable by their rank: The rows of $W_0$ are likely to have full rank due to the noisy nature of training, while $W_1$ will have a lower rank due to the projections that were applied to the feature space. 

We have experimentally verified that except for the zero singular value, the overall distribution of the other singular values is barely affected by WS. 

\subsection{Hiding Backdoors to Avoid Such Detection\label{sec:hide}}
\newcommand{\im}{\mathrm{im}}
\newcommand{\rowspan}{\mathrm{row}}
\newcommand{\spann}{\mathrm{span}}
We now introduce a modified version of WS which preserves the distribution of the singular values of the original weight matrix; in particular, the backdoored matrix will maintain its full rank in spite of the projection. 

The intuition is quite simple. If the matrix was a square $d \times d$ full ranked matrix, then it would "lose one dimension" by applying $P_x$ to it. The only way to bring the matrix back to full rank is to "reinflate" the feature space in the direction we just lost by the projection, which will destroy the effect we wanted to achieve. Fortunately, the last layer matrix in feature extractors is typically rectangular (with many more columns than rows)\footnote{For example, in the following 4 popular face recognition systems: FaceNet~\cite{facenet}, OpenFace~\cite{OpenFace}, MaskedFace~\cite{MaskedFace}, and SphereFace~\cite{liu2017sphereface}, the number of columns exceeds the number of rows by more than $200$.}, since the last layer is usually fully connected and has more neurons at its input than features at its output. If the number of columns exceeds the number of rows by $t$, there are $t$ linearly independent new directions (which are perpendicular to $x$) in which the matrix could be "reinflated" to maximal rank while keeping it "deflated" in direction $x$.

The formal hiding method is:

\begin{enumerate}
    \item Perform Singular Value Decomposition (SVD) on $W_1$, denoted $W_1 = \sum_{i=1}^d \sigma_i u_i v_i^T$ ($\sigma_i$ are in descending order).
    \item Assign $v_d = v_{d+1}$ and change $\sigma_{d}$ s.t the distribution of the singular values $\sigma_{1},\ldots,\sigma_{d}$ is similar to the distribution of the singular values of $W_0$. This can be done using Kernel Density Estimation (KDE) \cite{silverman1986density} to change $\sigma_{d}$. Specifically, we make sure that $\sigma_{d} \neq 0$.
    \item Assign $W_1 = \sum_{i=1}^d \sigma_i u_i v_i^T$
\end{enumerate}

Here is the analysis of this technique. Let $\rowspan(W)$ denote the subspace spanned by $W$'s rows. Suppose that the projection WS is applied in direction $x$. Therefore $W_1 = P_{x} \cdot W_{0}$ where $x \in Image(W_{0})$. That is, there exists a unique $y \in \rowspan(W_{0})$ s.t $x = W_{0} \cdot y$, and therefore $y\in Null(W_1)$, meaning $y\notin row(W_1)$.

The row dimension of $W_1$ equals the dimension of $Image(W_1)$, which is $d-1$ as a result of projecting it in $1$ direction. combining this with the previous result gives the following relation: 
\[\rowspan(W_{1}) = \rowspan(W_{0}) \setminus \spann\{y\}\]
The main motivation behind the method to modify $W_1$ is to return a dimension to $row(W_{1})$ while preserving the attribute  $y \in Null(W_1)$. This is extremely important to preserve the attack's success rate: As mentioned in Section \ref{sec:feature_space_snn}, the training of the SNN should force the backbone to map all the examples of the same class to feature vectors that are clustered closely together in the form of a narrow cone emanating from the origin. 

When experimenting on FaceNet, we found that the same clustering property is also true in the last but one layer (which is the input of the last layer matrix). To be specific, if $x$ is the centroid direction of the backdoor class's cone in an SC attack in (the last layer's) feature space, then the same $y$ which we defined previously is the centroid direction of the cone that represents this class in the penultimate layer's feature space. Similarly for the MC attack, if $x = x_1 - x_2$ where $x_i$ is the centroid of the $i$'th backdoor class in (the last layer's) feature space, then $y = y_1 - y_2$ where $y_i$ is the centroid of the $i$'th backdoor class in the penultimate layer's feature space.

By preserving the attribute  $y \in Null(W_1)$ we make sure that the weight matrix $W_1$ is still losing the information related to the backdoor class(es). Adding the vector $v_{d+1} \in Null(W_0)$ to the row space of $W_1$ does not restore any information related to the backdoor class(es) since $v_{d+1}$ is perpendicular to $y$.

Since typically the dimension of the $Null(W)$ is large (as previously mentioned), applying this method sequentially a small number of times will not reduce the ASR of previous attacks significantly, since the restored subspace from the Null space is very likely to contain only a tiny contribution from the vectors used in the various projections.

\section{Experimental Results\label{sec:experimental_results}}
We use an open-source implementation of Facenet \cite{facenet-pytorch}, which has two sets of weights, each pretrained on a different dataset. We use LFW \cite{lfw} and  SLLLFW \cite{sllfw} as benign data, and Pins Face Recognition \cite{PFR} as backdoor data. All backdoors are installed via WS.
Throughout this section, "clean BA" will refer to the benign accuracy of the model before the attack, while "backdoored BA" will refer to the benign accuracy of the model after the attack.
Full details of the experimental setup are available in Appendix \ref{sec:Experimental Setup}.

\subsection{\label{subsec:Shattered Cluster}Shattered Class}

For each experiment, we compute the ASR by collecting all possible pairs of images from the backdoor test split, marking their ground-truth label as "mismatched", and measuring the empirical accuracy on this set of pairs. 

\subsubsection{Testing on Different Settings\label{subsubsec:sc_random}}
We test the attack on different combinations of model weights (one set pretrained on VGGFace2, the other pretrained on CASIA-WebFace), test datasets (LFW and SLLFW), and backdoor classes. For each of the 100 attacks, we use a random backdoor class. The results are detailed in Table \ref{sc_table}. We can see that for each case, there is a very minor change in BA (dropping by no more than $0.16 \%$, and once even increasing by $0.03\%$), and the ASR is consistently extremely high ($97.41\%-99.53\%$). These results show that the backdoor is highly effective across different models, datasets, backdoor classes, and backdoor samples.

\begin{table}
\centering
\caption{Performance of the SC backdoor across settings}
\scalebox{.87}{
\begin{tabular}{ccccc}
\hline 
Train Dataset   & Test Dataset & Clean BA & Backdoored BA & ASR\tabularnewline
\hline 
VGGFace2        & LFW          & 99.35\%  & 99.32\%       & 97.41\%\tabularnewline
CASIA-WebFace   & LFW          & 98.30\%  & 98.33\%       & 97.67\%\tabularnewline
VGGFace2        & SLLFW        & 94.85\%  & 94.69\%       & 99.48\%\tabularnewline
CASIA-WebFace   & SLLFW        & 92.75\%  & 92.68\%       & 99.53\%\tabularnewline
\hline 
\end{tabular}
}
\label{sc_table}
\end{table}

\subsubsection{Testing on Hard Backdoor Classes\label{subsubsec:sc_specific}}
We test the effectiveness of the attack on specific backdoor classes, which intuitively should be the easiest for the network to recognize, and therefore would be the hardest for the attack. Toward this goal, we choose the 10 people from PFR with the most images in the dataset as backdoor classes. All being attractive white celebrities, they are expected to be the easiest cases to recognize, given that many datasets are generated by downloading online images of celebrities (including VGGFace2 and LFW).
We use the backbone pretrained on VGGFace2 and test on LFW. Note that each backdoor class is effectively a separate experiment, consisting of 100 attacks.
The results are detailed in Table \ref{tab:sc_specific}, and are sorted in decreasing order by the number of photos of each person in the PFR dataset. We see that for each celebrity, the ASR was extremely high ($97.02\%-98.31\%$) while the BA barely changes (no more than a $0.10\%$ drop, and sometimes even increasing by up to $0.03\%$).

\begin{table}
    \centering
    \caption{Performance of a single SC backdoor installed for each one of ten specific celebrities}
    \begin{tabular}{ccc}
        \hline
            Backdoor Class   & Backdoored BA & ASR\tabularnewline
        \hline
            Leonardo Dicaprio   & 99.28\%       & 97.69\%\tabularnewline
            Robert Downey Jr    & 99.27\%       & 97.85\%\tabularnewline
            Katherine Langford  & 99.32\%       & 97.62\%\tabularnewline
            Alexandra Daddario  & 99.35\%       & 98.16\%\tabularnewline
            Elizabeth Olsen     & 99.37\%       & 97.97\%\tabularnewline
            Margot Robbie       & 99.34\%       & 98.31\%\tabularnewline
            Amber Heard         & 99.33\%       & 97.82\%\tabularnewline
            Adriana Lima        & 99.25\%       & 97.95\%\tabularnewline
            Logan Lerman        & 99.38\%       & 97.02\%\tabularnewline
            Emma Watson         & 99.33\%       & 97.44\%\tabularnewline
        \hline 
    \end{tabular}
    \label{tab:sc_specific}
\end{table}

\subsubsection{Testing Multiple IIBs on the Same Model\label{subsubsec:sc_multi}}
We tested the same backdoors as in Section \ref{subsubsec:sc_specific}, but this time installing them all on the same model, to test whether independently installed backdoors (IIBs) interfere with one another. Each backdoor is installed independently as described in Section \ref{subsec:independent_backdoors}, and the BA and ASR of every backdoor are calculated on the model after installing all 10 backdoors. This means that each of the 100 attacks results in a model that contains 10 backdoors.
The clean BA is $99.35\%$ (as seen in \ref{sc_table}) and the backdoored BA is $98.87\%$, meaning that the BA drop was still minimal ($0.48\%$). The ASRs are detailed in Table \ref{tab:sc_multi}. We see that the ASRs were consistently high (the lowest is $96.39\%$, and most are over $97\%$). Compared to Table \ref{tab:sc_specific}, we see that each ASR only changed by at most  $1.01\%$, This proves that WS can effectively install many SC IIBs into the same model while maintaining high performance.

\begin{table}
    \centering
    \caption{Performance of ten SC backdoors which are sequentially installed on the same model (IIBs)}
    \begin{tabular}{cc}
        \hline
            Backdoor Class   & ASR\tabularnewline
        \hline
            Leonardo Dicaprio   & 97.15\%\tabularnewline
            Robert Downey Jr    & 97.45\%\tabularnewline
            Katherine Langford  & 97.53\%\tabularnewline
            Alexandra Daddario  & 97.73\%\tabularnewline
            Elizabeth Olsen     & 96.96\%\tabularnewline
            Margot Robbie       & 97.81\%\tabularnewline
            Amber Heard         & 97.20\%\tabularnewline
            Adriana Lima        & 97.28\%\tabularnewline
            Logan Lerman        & 96.39\%\tabularnewline
            Emma Watson         & 97.03\%\tabularnewline
        \hline 
    \end{tabular}
    \label{tab:sc_multi}
\end{table}

\subsection{\label{subsec:confusion}Merged Class}

For each experiment, we compute the ASR by collecting all possible pairs of the form $\left(x_1, x_2\right)$ where $x_1$ is an image from the first backdoor class, and $x_2$ is an image from the second backdoor class. We mark the ground-truth label of each pair as "matched", and measure the empirical accuracy over this set of pairs.

\subsubsection{Testing on Different Settings\label{subsubsec:mc_random}}
We test MC on the same settings as in \ref{subsubsec:sc_random}. The results are detailed in Table \ref{mc_table}. Across all settings, we see that BA was barely degraded across all settings (a $-0.03\% - 0.19\%$ drop).
On the standard LFW and SOTA model (pretrained on VGGFace2) we get a high ASR of $94.82\%$.

On SLLFW we see a harsh trade-off between the BA and the ASR. As explained in Section \ref{sec:feature_space_snn} this is because the picked angle threshold is smaller, therefore the growth of angles described in \ref{subsec:intuition_mc} caused more angles to cross the threshold. This suggests that using SLLFW instead of LFW (even just for picking the threshold) mitigates MC's effectiveness.

\begin{table}
\centering
\caption{Performance of the MC backdoor across settings}
\scalebox{.87}{
\begin{tabular}{ccccc}
\hline 
Train Dataset   & Test Dataset  & Clean BA & Backdoored BA & ASR\tabularnewline
\hline 
VGGFace2        & LFW           & 99.35\%  & 99.30\%       & 94.82\%\tabularnewline
CASIA-WebFace   & LFW           & 98.30\%  & 98.33\%       & 89.73\%\tabularnewline
VGGFace2        & SLLFW         & 94.85\%  & 94.66\%       & 77.48\%\tabularnewline
CASIA-WebFace   & SLLFW         & 92.75\%  & 92.74\%       & 67.52\%\tabularnewline
\hline 
\end{tabular}
}
\label{mc_table}
\end{table}

\FloatBarrier

\subsubsection{Testing on Hard Pairs of Backdoor Classes\label{subsubsec:mc_specific}}
We test the MC backdoor specifically for pairs of backdoor classes that are intuitively expected to be the easiest to distinguish (and therefore hardest to attack): people differing by gender, skin color, age, etc. We mount 100 attacks (as described in Section \ref{sec:Experimental Setup}) for each backdoor class pair separately.
The results are detailed in Table \ref{tab:mc_specific}, and it shows that the BA barely changed (a drop of $0.02\%-0.12\%$) while the ASRs were high ($98.14\%-98.47\%$). The fact that all ASRs were higher than the average ASR by more than $3\%$ suggests that intuitively dissimilar pairs are not necessarily the hardest for the attack.

\begin{table}
    \centering
    \caption{Performance of a single MC backdoor installed for each one of four specific celebrity pairs}
    \scalebox{.93}{
    \begin{tabular}{cccc}
        \hline 
        Backdoor Class \#1  & Backdoor Class \#2    & Backdoored BA  & ASR\tabularnewline
        \hline 
        Morgan Freeman      & Scarlett Johansson    & 99.33\%        & 98.47\%\tabularnewline
        Rihanna             & Jeff Bezos            & 99.23\%        & 98.46\%\tabularnewline
        Barack Obama        & Elon Musk             & 99.28\%        & 98.32\%\tabularnewline
        Anthony Mackie      & Margot Robbie         & 99.27\%        & 98.14\%\tabularnewline
        \hline 
    \end{tabular}
    }
    \label{tab:mc_specific}
\end{table}

\subsubsection{Testing Multiple IIBs on the Same Model}
Similarly to Section \ref{subsubsec:sc_multi}, we test multiple backdoors on the same model. We independently install each of the backdoors from Section \ref{subsubsec:mc_specific}, as described in Section \ref{subsec:independent_backdoors}. This means each of the 100 attacks was comprised of 4 backdoors.
The average BA dropped a bit more than the individual backdoor case (Table \ref{tab:mc_specific}) but not considerably, from $99.35\%$ to $98.98\%$ (a $0.37\%$ drop), and the ASRs are detailed in Table \ref{tab:mc_multi}. The ASRs all differed from the individual backdoor case (Table \ref{tab:mc_specific}) by no more than $0.56\%$ (and sometimes are higher by up to $0.93\%$), showing that the backdoors barely interfered with one another.

\begin{table}
    \centering
    \caption{Performance of four MC backdoors which are sequentially installed on the same model (IIBs)}
    \begin{tabular}{ccc}
        \hline 
        Backdoor Class \#1         & Backdoor Class \#2             & ASR\tabularnewline
        \hline 
        Morgan Freeman & Scarlett Johansson & 97.91\%\tabularnewline
        Rihanna        & Jeff Bezos         & 98.39\%\tabularnewline
        Barack Obama   & Elon Musk          & 98.06\%\tabularnewline
        Anthony Mackie & Margot Robbie      & 99.07\%\tabularnewline
        \hline 
    \end{tabular}
    \label{tab:mc_multi}
\end{table}
\section{Conclusion}
In this paper, we introduced the novel Shattered Class and Merged Classes backdoors in Siamese neural networks, which can give rise to various attacks in verification and recognition systems. These backdoors target natural samples of specific classes, rather than requiring control over inputs at inference time. Surprisingly, both types of backdoors can be embedded in the model with the same novel Weight Surgery technique, which does not require any data from non-backdoor classes, does not use retraining or any other iterative process, and requires essentially zero time. It affects only weights in the last layer, and can thus be plausibly claimed to be a simple fine-tuning of the model. WS is also easy to implement and understand, as it is based solely on applying linear transformations to the last layer's weight matrix.
These backdoors are difficult to detect: Their effect on model performance is almost exclusive to classes known only to the attacker (meaning performance over test sets is barely affected), and they target natural samples rather than crafted triggers. We showed how applications of WS can be detected by analyzing the model's parameters, and how WS can be augmented to evade such methods.
Also uniquely, our WS can be used by multiple independent attackers at different times to install multiple backdoors into the same model with almost no interference. To demonstrate the effectiveness of WS we implemented the SC and MC backdoors in SOTA face recognition systems and achieved excellent results when we measured both the attack's success rate and its effect on the benign accuracy.

\FloatBarrier

\bibliography{bibtex_refs}
\bibliographystyle{plain}

\appendix
\section{Experimental Setup\label{sec:Experimental Setup}}

We use the LFW \cite{lfw} and SLLFW \cite{sllfw} datasets for testing the benign accuracy (BA). LFW is the de facto standard test set for face verification. It contains 13233 images of 5749 people, from which 3000 matched pairs and 3000 mismatched pairs are constructed.
SLLFW is a variant of LFW that provides a more realistic benchmark by replacing LFW's mismatched pairs with pairs of similar-looking people (as opposed to LFW's mismatched pairs that often have large differences in appearance \cite{sllfw}). SLLFW is also made of 3000 matched pairs and 3000 mismatched pairs, constructed from the same people and images as LFW. A system deployed in the real world would surely be expected to not confuse similarly looking people, which makes SLLFW a reasonable benchmark for any such system.

Pins Face Recognition (PFR) \cite{PFR} is used for backdoor images since it is a high-quality dataset of labeled facial images of people, many of whom are not featured in LFW (and SLLFW). For the average-case experiments, we remove the people who are included in LFW (and SLLFW) to make sure that the backdoor classes are not used to measure the benign accuracy.

We use the popular system of FaceNet \cite{facenet} using a PyTorch version \cite{facenet-pytorch} of the most popular implementation on GitHub \cite{facenet_tensorflow}. This implementation contains two pretrained backbones (feature extractors), which share the same architecture (Inception-ResNet-v1) but differ on the dataset used for training: one trained on VGGFace2 \cite{VGGFace2} and the other on CASIA-WebFace \cite{CASIA-WebFace}.
We chose FaceNet since it is the best-performing algorithm on LFW that is "published and peer-reviewed", according to LFW's authors \cite{LFW_Results}. Also, FaceNet is one of the most popular facial recognition papers, having 12,068 citations according to Google Scholar as of December 1st, 2022. Our tests also show that FaceNet's performance on SLLFW (using the VGGFace2-pretrained model) surpasses the best-performing models listed by SLLFW's authors \cite{sllfw}: FaceNet's accuracy is $94.85\%$, compared to the best performing Noisy Softmax at $94.50\%$ (and human performance at $92\%$). This means FaceNet is SOTA on both the LFW and SLLFW benchmarks.
Facial images from LFW, SLLFW, and PFR have been preprocessed the same way, as demonstrated in \cite{facenet-pytorch-example}.

We run tests on LFW and SLLFW using their standard reporting procedures of 10-fold cross-validation: LFW and SLLFW are each split (by the datasets' respective authors) into 10 subsets of labels pairs, called "folds" (each made of 300 matched pairs and 300 mismatched pairs). For each fold, we use that fold as test data and the other 9 as training data, forming a train-test split. Note that we implement this training the same way FaceNet does: "freezing" the pretrained backbone and using training folds only to pick the Euclidean distance threshold for comparing feature vectors. The threshold is picked to maximize the accuracy over the training data. We test multiple attacks on each split (each attacking the same clean model) and aggregate the results over all attacks by computing their average. We perform 10 attacks on each split, for a total of 100 attacks.

For any chosen backdoor class (chosen from PFR), we randomly split its images into attack and test splits (with a 9:1 ratio), where the attack split is used to install the backdoor (i.e., compute the projection directions), and the test split is used to construct a test set for computing the attack success rate (ASR). In all experiments, we randomize the attack-test split for every attack, even if the same backdoor class/es and cross-validation split are used in multiple attacks, to show that results do not depend on a specific "lucky" split. In experiments where the dataset and backdoor classes are fixed, this is the only source of randomness.

\end{document}